\documentclass[12pt,preprint]{aastex}

\shorttitle{Globular Cluster and Halo Formation}
\shortauthors{Li \& Burstein}

\begin{document}

\title{Nitrogen Overabundance: Globular Cluster and Halo Formation}

\author{Yong Li\altaffilmark{1},
David Burstein\altaffilmark{1}}

\altaffiltext{1}{Department of Physics and Astronomy, Box 871504, Arizona
    State University, Tempe, AZ  85287-1504}

\begin{abstract}

Halo globular clusters pose four succinct issues that must be solved
in any scenario of their formation: single-age, single metallicity
stellar populations; a lower limit ([Fe/H]$\sim$-2.3) to their average
metallicity; comprising only 1\% of the stellar halo mass, and being
among the oldest stars in our Galaxy. New spectra are presented of
Galactic stars and integrated spectra of Galactic globular clusters
which extend to 3250$\rm \AA$. These spectra show show that the most
metal-poor and among the best-studied Galactic globular clusters show
strong NH3360 absorption, even though their spectral energy
distributions in the near-UV are dominated by blue horizontal branch,
AF-type stars.  These strong NH features must be coming from the
main sequence stars in these clusters. These new data are combined
with existing data on the wide range of carbon and nitrogen abundance
in very metal-poor ([Fe/H] $<$ -3.5) halo giant and dwarf stars,
together with recent models of zero-metal star formation, to make a
strawman scenario for globular cluster formation that can reproduce
three of the above four issues, and well as related two of the three
issues pertaining to nitrogen overabundance. This strawman proposal 
makes observational and theoretical predictions that are testable, 
needing specific help from the modelers to understand all of the 
elemental constraints on globular cluster and halo formation.

\end{abstract}


\section{Introduction}

Any theory of globular cluster (GC) formation must be able to explain
the following observations: 1. The low mass stars we see today in halo
GCs comprise a single-age, single [Fe/H] set of stars, with specific
abundance variations among proton-capture elements
\citep[e.g.,][]{grat01}. The gas from which these stars formed was not
blown away from the winds/supernovae from the more massive stars in
these clusters. 2. There appears to be a lower limit to the average
[Fe/H] of a GC, being near [Fe/H] = -2.5 for the most metal-poor
clusters in our Galaxy and in other galaxies. 3. GCs comprise about
1\% of stellar mass of the halo of our Galaxy.  4. The halo clusters
are the oldest stars in our Galaxy, likely formed during the first
burst of star formation.  Given the age now put on the Universe by the
WMAP observations, 13.7 Gyr \citep{sper03}, and the age estimates for
metal-poor GCs ($\sim 13$ Gyr) \citep[e.g.,][]{grun00}, it is clear
that the halo, most metal-poor GCs (e.g., M92, M15) formed very 
shortly after recombination.

Recent observations \citep[e.g.,][]{nrb01,nor02,aoki02,c02} have found
that a few halo subgiant and giant stars with [Fe/H] = -3.5 to -5.3
have surprisingly high nitrogen and carbon abundances. Such metal-poor
stars are almost certainly formed coevally with the halo GCs. 
Aoki et al. find two of their stars (CS 22949-037 and CS29498-043) 
have spectroscopic evidence of [N/Fe] = 2.3 $\pm$ 0.4, similar to 
what was found by \citet{bn82} for two other stars.

\citet{bur84} and \citet{pbo98} have found both Galactic GCs and M31
GCs to have enhanced CN features (near 4150$\rm \AA$ and 3880$\rm
\AA$) in their integrated spectra.  Ponder et al. also show that the
NH molecular band at 3360$\rm \AA$ is very strong in the integrated
spectra of four M31 GCs, including 3 of the ones studied by Burstein et al.

YL is in the process of completing a Ph.D. thesis, in which he has
obtained spectrophotometry from 3250$\rm \AA$ to 1$\mu$ at 10$\rm \AA$
spectral resolution for 125 stars with IUE data, as well as for 8
Galactic GCs. \S~2 presents our spectra for these 8 Galactic GCs,
together with spectra for the 4 M31 GCs studied in Ponder et al. and
the NH, CH, Mg$_2$, and [Fe/H] data for these GCs and graphically for
NH, CH and [Fe/H] for the stars in YL's thesis. \S~3 combines our
results with those for the metal-poor halo stars and presents a
strawman scenario for the formation of the oldest stars in our Galaxy
which can fit most of the observational issues given above, plus some
of those that arise with carbon and nitrogen overabundance in the
oldest stars.  Equally important, this scenario makes several
predictions which can be observationally and theoretically tested.

\section{Near-UV Observations of Galactic and M31 Globular Clusters,
and Galactic Stars}

YL's observations of Galactic GCs and stars were made with the Boller
\& Chivens CCD spectrograph on the 2.3m Bok telescope of the
Steward Observatory.  Li integrated the light from each GC with a 4$'$
spectrograph slit to synthesize their spectra over a region
approximately $3' \times 4'$ in size.  Data for all stars and Galactic
GC are fluxes and cover the wavelength range 3250$\rm \AA$ to 1$\mu$
at 10-12$\rm \AA$ spectral resolution.  Figure~1 presents the
integrated spectra from $\rm 3250 \AA$ to 4900$\rm \AA$ that Li has
obtained for 8 Galactic GCs, ranging from M92 to M71 in metallicity.  
The high signal-to-noise (S/N) of YL's GC observations is evident in 
the high degree of similarity of the spectra of M92 and M15. Also 
presented In Figure~1 for completeness are the analogous integrated 
spectra obtained with the HST/FOS for 4 M31 GCs by Ponder et al.

\citet{tl84} pointed out that NH and CH (the G-band) have essentially
the same dissociation energy (a difference of only 0.05 eV), so that a
relative comparison of carbon to nitrogen abundance can be made in
individual stars by comparing the strength of the molecular bands of
these two molecules. \citet{dc94} and YL's stellar observations show
that NH, like CH, is absent in Galactic stars earlier than F2 or so, 
and is stronger in giant stars than in dwarf stars.

It is clear from these Figure 1 that the NH molecular band is present
in the spectra of all 12 GCs, and is especially strong in the more
metal-rich M31 GCs (Mayall II, K280 and K58). What is even more
remarkable are the observed NH strengths in M92, M15, M53, among the
most metal-poor GCs in our Galaxy: Inversion of the Ca II H\&K lines
(with H being stronger than K) is indicative of domination of A-F
stars near 4000$\rm \AA$ (likely due to the blue horizontal branches
of these clusters; e.g, \citet{pio02}), and is seen in the 7 more
metal-poor Galactic GC spectra (but not for one metal-rich GC in
our sample, M71).  Along with early-type Galactic stars, metal-poor
stars earlier than F-type do not have NH in their spectra.  Moreover,
stellar population models indicate that giant stars contribute at most
15\% of the flux at 3360$\rm \AA$ (R. Peterson, private
communication). The enhanced NH features seen in some giant
stars in metal-poor GCs \citep[e.g.,][]{s81,ketal82,tetal83,ssbsk99}
cannot alone produce the strong NH we see.  These strong NH features
seen in integrated light must be mostly due to the main sequence stars
in these GCs.

In Figure~2 we plot the absorption line strengths of CH (G-band)
vs. NH, and CH and NH vs. [Fe/H] for the stars in Li's thesis,
together with those of the 12 GCs shown in Figure~1.  The CH data for
Galactic and for M31 GCs are taken from \citep{tr98}. [Fe/H] for
Galactic GC are taken from the online data given by Harris
\citep[e.g.,][]{har96}; for stars from the compilation in
\citet{wor94}.. The definition given by \citet{dc94} is used for the
NH feature.

Five issues are evident from Figure~2: 1. The CH measures for the all
of the Galactic and M31 GC place their indices mostly among the
Galactic dwarf stars. 2. NH is clearly enhanced relative to CH, [Fe/H]
(and also Mg$_2$, not shown) for the 8 most metal-poor GCs, relative
to that observed in Galactic stars.  3. The strength of NH in the MS
stars in the metal-poor clusters has to be stronger than what we see
in their integrated light, as the flux in the near-UV in these
clusters is dominated by their blue horizontal branch stars. Yet, if we
examine the horizontal branch ratio for these clusters
\citep[e.g.,][]{har96}, we do not find that it correlates with NH
strength for the 7 metal-poor Galactic GC.  Perhaps there is range of
nitrogen abundance among the metal-poor Galactic GC?  4. As a group,
NH is far stronger in the 3 more metal-rich M31 GCs than it is in the
Galactic GCs at a given value of CH or [Fe/H].  5. There are three
well-known nitrogen-rich stars (HD 122563, HD 165195, and HD 201626,
e.g, \citet{sned74,yor83,ssbsk99}) that have nitrogen abundances as
high as those in the {\it integrated} spectra of the more metal-rich
M31 GCs. However, as giant stars do not dominate in the near-UV in
these GCs, the mixing that produces the strong NH in these Galactic
giant stars cannot be the sole source of the strong NH feature in the
M31 GC.
 
A full analysis of the nitrogen overabundance in these Galactic GCs
will require obtaining high S/N spectra of their lower main sequence
stars, a task that still remains to be done. However, given that the
structure of the NH feature in these GCs is similar to its structure
seen in metal-rich stars (like the Sun; e.g., \citet{nor02}), we are
relatively safe in predicting that for the M31 and Galactic GCs,
[N/Fe] ranges from 0.7 to 2.5.  For example, {\it if} [N/Fe] = 1.7 in
M92 stars, then the N abundance in M92 stars means that the total mass
of nitrogen, assuming $1-2 \times 10^5$ M$_\odot$ of stars, is $\rm
30-50 M_\odot$.

\section{A Strawman Formation Scenario}

We make the reasonable initial assumption that the overabundance of
nitrogen seen in a few halo giant stars with [Fe/H]$<$-3.5, and in the
integrated spectra of the most metal-poor Galactic and M31 GCs is 
primordial in origin.  There is simply too much nitrogen
overabundance for any secondary enhancement of nitrogen to take place
via 5-7 M$\odot$ asymptotic giant branch stars in these clusters
\citep[such as advocated by ][]{lslp02}, without appealing to unusual
initial mass functions.  Hence, to the list of observations that any
formation scenario for GCs must fit, given in \S~1, we now can add
three more: 5. Produce a marked overabundance of nitrogen in the most
metal-poor, oldest GCs.  6. Produce a similar marked overabundance of
nitrogen in a {\it subset} of the oldest, most metal-poor halo
stars. 7. Produce an intrinsic range of nitrogen abundances between
the GCs among different galaxies, and perhaps also among the oldest
GCs within a given galaxy.

Our strawman proposal can, in broad-brush strokes, can currently 
fit 5 of these 7 issues for GCs and halo stars, and can also be 
consistent with what we know about the early Universe.  Cayrel 
\citep{cay86,cay87} outlined much of this scenario; here we expand 
on his scenario, given that we have more information than was 
available in the 1980's.

We begin with the idea that formation of GCs most likely
took place with zero-metal, H+He primordial gas.  Then, take into
account that the Jean's mass of this zero-metal gas is $10^5 -
10^6$ M$_\odot$ for a long time after recombination \citep[e.g.][and
references therein]{rfb88}.  Now insert the issue that \citet{nor02}
brought up, namely, the kind of star can produce such remarkable
overabundances of C and N in the oldest stars, that the zero-metal 
stellar models of \citet{fwh01} seem to fit the CNO issue. These are 
200-500 M$_\odot$ zero-metal rotating stars which quickly burn some 
of their hydrogen to helium, then helium to carbon, then undergo 
carbon burning of hydrogen. The net effect of this is that when this 
supermassive star goes hypernova, it expels out many solar masses of 
CNO product.  

Now suppose that modelers can produce a hypernova from zero-metal gas
that produces not just C, N and O but also the various other proton
capture reactions needed to produce the range of elemental abundances
differences seen in Galactic GC \citep[e.g.,][]{grat01}.  A tall order
to fill, but one that needs to be explored. And also suppose that
these new models show that these hypernova produce a range of
abundance of CNO, while producing a set amount of higher order metals.
Yet another tall order.  If each hypernova produces 20-40 solar mass
of nitrogen, then perhaps just one hypernova in the centers of
$10^5-10^6$ M$_\odot$ Jean's mass clumps is needed to produce the
nitrogen overabundance we see in the most metal-poor GCs.  The most
likely place for such zero-metal stars to form is at or near the
centers of the zero-metal Jean's masses that have formed in the early
Universe. And, most of those masses will be found within the dark
matter confines of what will eventually become a giant galaxy.

In the case of the GCs, these massive stars sit very close
to the centers of these Jean's mass clumps, and their hypernovae go
off symmetrically more or less simultaneously seeding the rest of the
mass of the clump with its pollutants, and uniformly crunching the gas
to make stars out of this gas. The resulting stars will then have a
base level of metallicity, and a large overabundance of C and/or N
and/or O, of which we clearly see today the N overabundances.  As these
clumps are close together at early epochs in the Universe, merging of
clumps is also likely, producing clusters with bimodal CNO abundances
\citep[e.g.,][]{hsg03}.

On the other hand, the likelihood that such symmetry exists in the
hypernovae explosions that will occur near the centers of these $\rm
10^6 M_\odot$ zero-metal clumps is small.  In fact, given that
globular cluster stars comprise only 1\% of the halo stars, it is
likely a 1\% probability.  In almost all of the cases, therefore, the
hypernovae will go off {\it asymmetrically}, crunching just some of
the remaining gas, but releasing the rest of its energy (and thereby,
if this is true for all galaxies, re-ionizing the Universe).  This
will result in most halo stars getting different portions of the
hypernova elemental products. This, then, can produce the intrinsic
range of C and N overabundance that is observed in some, but not all,
metal-poor halo stars.  It is also likely that formation of this kind
also existed in the masses that formed the centers of giant elliptical
galaxies, as \citet{tr00a,tr00b} find that N is overabundant in their
centers, along with $\alpha$-product elements.

This scenario can readily reproduce five of the seven issues
discussed above for GC formation.  What it cannot reproduce is the
apparent systematic difference in nitrogen abundance seen in the
present data between the M31 GCs and the Galactic GCs.  Nor can it
yet reproduce the elemental abundance similarities and differences
found among Galactic globular cluster stars versus those in the halo.
Rather than take the difference between M31 GC and Galactic GC at 
face value, we note that these 4 M31 GCs are more luminous that 
even $\omega$ Cen in our own Galaxy.  NH observations of less luminous 
M31 globular clusters will be needed, together with observations of 
many other Galactic GCs and GCs in other galaxies (e.g, the Fornax 
dwarf galaxy, Large Magellanic Cloud, Cen A, M81), before we can 
understand if there are true systematic differences in nitrogen 
abundance among the oldest GCs in different galaxies.

Our scenario for GC and halo star formation makes five testable
predictions: 

1. Strong NH features should exist in other, metal-poor halo GCs in
our Galaxy as well as metal-poor (and old metal-rich) GCs in other
galaxies. It remains to be seen how much variation we see among the
GCs within a given galaxy, and from galaxy-to-galaxy.

2. The hypernovae that produce the GCs might produce black
holes. Depending on how that black hole is fed (and it might be very
starved within the beehive that are GCs), its mass might be only
50-100 M$_\odot$, making it very hard to detect in most GCs. In this
regard, we note the suggestion that large, massive (2000-20,000
M$_\odot$) blackholes are in two of the GC for which we show strong NH
features: Mayall II (MII in Figure~1) in M31 \citep{vdm02} and in the
metal-poor Galactic GC, M15 \citep{grh02}, although these
interpretations are being disputed \citep{baum03a,baum03b}.

3. Those GCs that form later in time will not produce such hypernovae,
so will not likely show the kind of nitrogen overabundance relative to
iron that is seen in the metal-poor globular clusters or the most
metal-poor galactic stars.  How much younger should these other GCs
be? If we say that the oldest GCs are 13 Gyr old, then ones that are,
say, 8-9 Gyr old might fit this bill.  Hence, much younger GCs should
{\it not} show NH enhancements. The mild oxygen overabundance seen in
[Fe/H] $>$ -1.5 stars \citep[e.g.][]{edv93} is consistent with this
issue. We note that the data presented here for the younger disk GC
M71, compared to that of its older, more metal-poor cousins, is the
most consistent with this prediction among the Galactic GCs.

4. The most metal-poor halo stars should show a continuum of CNO
abundances.

5. Zero-metal star models can be made that reproduce the various
proton-capture abundance issues that are now known to exist among
Galactic GC stars, while preserving their small range in overall
[Fe/H] as well as intrinsic N abundance differences among
different galaxies. Altogether, a significant challenge for the
modelers.

Exploring the observational and theoretical consequences of this halo
and GC formation scenario will take the efforts of both observers
(near-UV integrated spectra of GCs in our Galaxy and in other
galaxies; similar spectra of individual main sequence GC and halo
stars in our Galaxy) and theoreticians (to interpret these spectra and
to make new zero-metal gas stellar models). We put forward this
strawman scenario with the strong likelihood that much of it will be
have to be modified in the future, if not completely changed. However,
we have to start somewhere to understand the now clearly-defined 
nitrogen-based issues in old stars, and here is where we choose to start.

\acknowledgements

We thank Donald Lynden-Bell for very stimulating conversations
about globular cluster formation scenarios, and John Norris, Mike
Bessell and Ken Freeman for helpful suggestions.  We also thank
the anonymous referee for helpful comments for reshaping this paper.

\clearpage

\begin{figure}
\epsscale{0.8}
\plotone{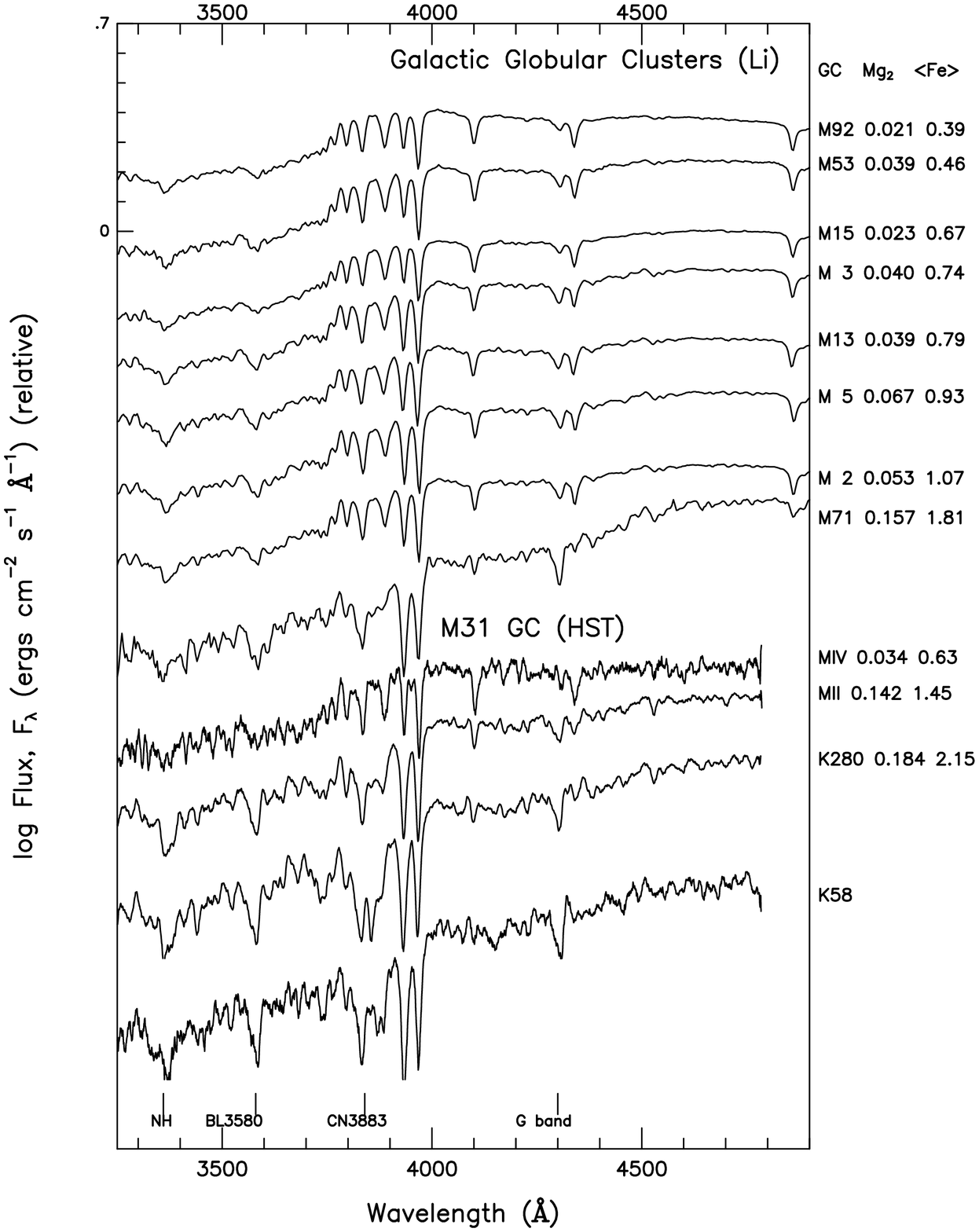}
\figcaption{Integrated spectra from 3250$\rm \AA$ to 4900$\rm \AA$ of
8 Galactic globular clusters from Li's Ph.D. thesis (top) and from
3250$\rm \AA$ to 4800$\rm \AA$ for 4 M31 globular clusters from Ponder
et al. (bottom), plotted as log flux vs. wavelength.  Each spectrum is
plotted on a log scale of 0.7 dex of flux.  The Messier numbers of the
Galactic clusters are given.  MII and MIV are Mayall II and Mayall IV,
280 and 58 are the Vitesnik numbers for these M31 GCs (cf. Ponder et
al. 1998). Also given are the Mg$_2$ and $\rm <Fe>$ measures for 11 of
these clusters from Trager et al. (1998).}
\label{fig1}
\end{figure}

\begin{figure}
\epsscale{0.55}
\plotone{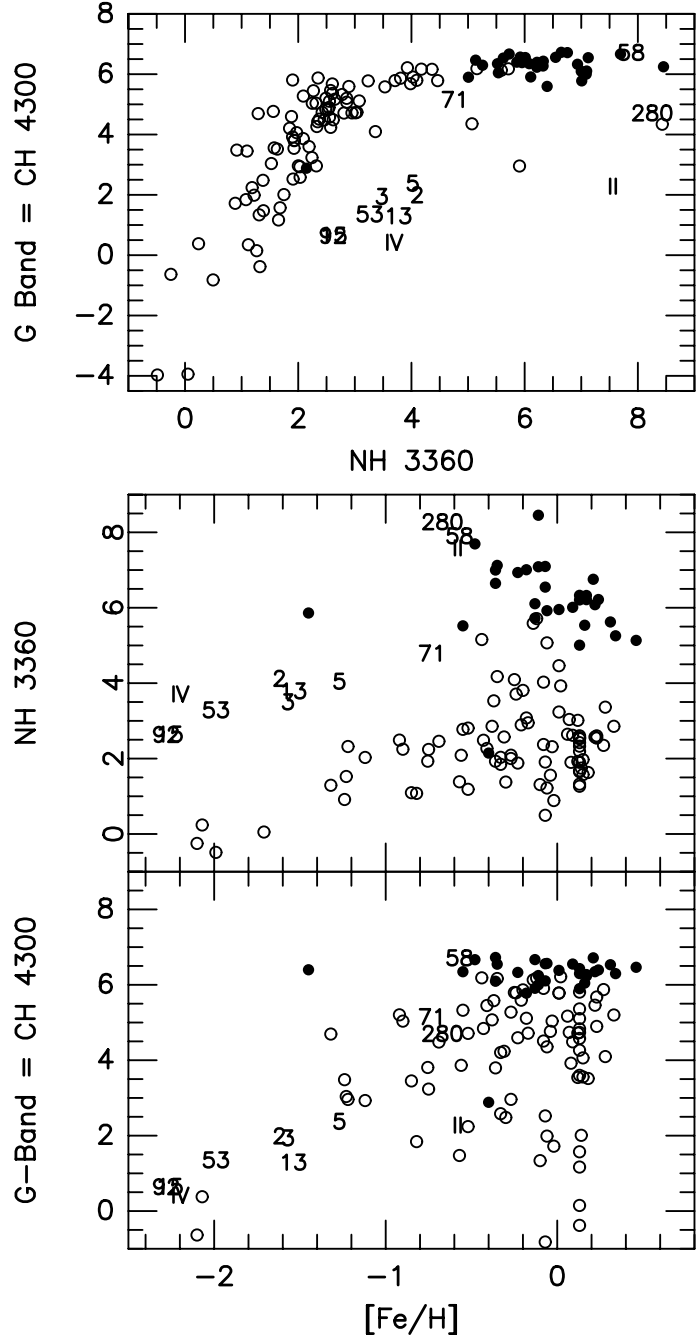}
\figcaption{(top) The absorption line measure of CH (G-band) plotted
versus NH for Galactic dwarf stars (open circles), Galactic giant
stars (closed circles), the Galactic GCs (given by their Messier
numbers) and M31 GCs (given by their names as listed in Figure~1).
The data for the 125 Galactic stars comes from YL's thesis.
(lower two graphs) NH vs. [Fe/H] and CH vs. [Fe/H].  The NH and CH
values for the Galactic GC are, in order of Messier number, CH and
NH: (2: 1.966,4.099), (3: 1.912,3.476), (5: 2.351,4.022),
(13: 1.268,3.766), (15: 0.626,2.621), (53: 1.328,3.262), 
(71: 5.127,4.767), (92: 0.618,2.612).  The strong agreement of the
CH and NH values for M15 and M92 indicate the high S/N of our
GC spectra.}
\label{fig2}
\end{figure}

\end{document}